\documentclass[fleqn,usenatbib]{mnras}

\usepackage[T1]{fontenc}

\DeclareRobustCommand{\VAN}[3]{#2}
\let\VANthebibliography\thebibliography
\def\thebibliography{\DeclareRobustCommand{\VAN}[3]{##3}\VANthebibliography}

\usepackage{graphicx}	
\usepackage{amsmath}	
\usepackage{amssymb}	
\usepackage{subfigure}	
\usepackage{array}
\usepackage{booktabs} 
\usepackage[usenames,dvipsnames]{color}
\usepackage{colortbl}
\usepackage{color}
\usepackage{soul}
\definecolor{mygray}{gray}{.9}
\definecolor{mypink}{gray}{.5}
\usepackage{romannum}
\usepackage{newtxtext,newtxmath}

\makeatletter
\newlength{\abovecaptionskip}%
\makeatother
\usepackage{threeparttable}

\title[Reflection of MAXI J1535-571]{Analysis of the reflection spectra of MAXI J1535-571 in the hard and intermediate states}

\author[Yanting Dong et al.]{
Yanting Dong,$^{1}$\thanks{E-mail: ytd@zju.edu.cn}
Zhu Liu,$^{2}$
Youli Tuo,$^{3}$
James F. Steiner,$^{4}$
Mingyu Ge,$^{3}$
Javier A. Garc\'ia,$^{5,6}$
Xinwu Cao$^{1,7}$\thanks{E-mail: xwcao@zju.edu.cn}
\\
$^{1}$Institute for Astronomy, School of Physics, Zhejiang University, 38 Zheda Road, Hangzhou 310027, People’s Republic of China\\
$^{2}$Max Planck Institute for Extraterrestrial Physics, Giessenbachstrasse 1, 85748, Garching, Germany\\
$^{3}$Key Laboratory of Particle Astrophysics, Institute of High Energy Physics, Chinese Academy of Sciences, Beijing, People’s Republic of China\\
$^{4}$Harvard-Smithsonian Center for Astrophysics, 60 Garden St. Cambridge, MA 02138, USA\\
$^{5}$Cahill Centre for Astronomy and Astrophysics, California Institute of Technology, Pasadena, CA 91125, USA\\
$^{6}$Dr. Karl Remeis-Observatory and Erlangen Centre for Astroparticle Physics, Sternwartstr.~7, 96049 Bamberg, Germany
\\
$^{7}$Shanghai Astronomical Observatory, Chinese Academy of Sciences, 80 Nandan Road, Shanghai, 200030, People’s Republic of China\\
}

\date{Accepted XXX. Received YYY; in original form ZZZ}

\pubyear{2022}

\begin{document}
\label{firstpage}
\pagerange{\pageref{firstpage}--\pageref{lastpage}}
\maketitle

\begin{abstract}
We report results on the joint-fit of the \emph{NuSTAR} and \emph{HXMT} data for the black hole X-ray binary candidate MAXI J1535-571. The observations were obtained in 2017 when the source evolved through the hard, hard-intermediate and soft-intermediate states over the rising phase of the outburst. After subtracting continuum components, X-ray reflection signatures are clearly showed in those observations. By modeling the relativistic reflection in detail, we find that the inner radius $R_{\rm{in}}$ is relatively stable with $R_{\rm{in}}\lesssim 1.55 R_{\rm{g}}$ during the three states, which implies that the inner radius likely extends to the innermost stable circular orbit even in the bright hard state. When adopting $R_{\rm{in}} = R_{\rm{ISCO}}$, the spin parameter is constrained to be $0.985_{-0.004}^{+0.002}$ at 90\% confidence (statistical only). The best-fitting results reveal that the inclination of the inner accretion disc is $\sim70-74$ degrees, which notably conflicts with the apparent orientation of the ballistic jet ($\leqslant$45 degrees). In addition, both the photon index and the electron temperature increase during the transition from hard to soft state. It seems that the corona evolves from dense low-temperature in the LHS to tenuous high-temperature after the state transition, which indicates that the state transition is accompanied by the evolution of the coronal properties.

\end{abstract}




\section{Introduction}
\label{sec:intro}

During a typical outburst of a transient black hole binary, the black hole binary goes through different spectral states with spectral and timing properties changes. As the source luminosity increases, it evolves from the low/hard state (LHS) to the hard and soft intermediate states (HIMS, SIMS), then enters into the high/soft state (HSS, \citealt{bel2005, rem2006}). It is generally agreed that the change of spectral states is induced by the evolution of the accretion geometry of the black hole binary system. 

In the HSS, the source spectrum is dominated by thermal emission ($\sim 1$ keV) accompanied by a hard power-law tail. The accretion flow is composed of an optically thick and geometrically thin accretion disc \citep{ss1973} with its inner radius likely at the innermost stable circular orbit (\uppercase{isco}, \citealt{tan1995, ste2010}). In the LHS, the spectrum is dominated by the hard powerlaw X-rays, together with the very faint thermal component detected sometimes. The hard X-rays are produced by the inverse Compton scattering of thermal emission in a region of hot plasma, the so-called corona, and can be well described by powerlaw with $\Gamma \sim 1.4-2.1$. In the model of \citet{esi1997}, at a low accretion rate, the disc is truncated before it reaches the \uppercase{Isco}, and an advection-dominated accretion flow, which is evaporated from the accretion disc \citep{mey2000, liu2002, qia2010}, is in the inner region. A disc with a truncated inner radius of several tens to hundreds of $R_{\rm{g}}$ (the gravitational radius and calculated by $R_{\rm{g}} = GM/c^2$) has indeed been inferred by modelling the disc component of some BHXRBs, e.g.,  XTE J1118+480 \citep{esi2001}. With this truncated model, the transition from the LHS to the HSS can be well explained by the extending of inner radius down to the \uppercase{Isco} \citep{pla2014}. This model has also been invoked to explain the positive correlation between the X-ray photon index and the reflection strength \citep{zdz1999, ezh2020, pan2020}. 

On the other hand, a black hole binary will experience HIMS and SIMS before it enters the HSS. The thermal emission and the hard X-ray emission are both strong, which leads to a softer spectrum than that in LHS. The HIMS-SIMS transition can be very rapid. They are normally distinguished by the differences in their timing properties. For instance, either a type A or a type B QPO appears in the SIMS, while a type C QPO is often shown in the HIMS. Black hole binary also shows very weak variability in SIMS \citep{bel2005, bel2010}. These intermediate states, as the transition states between the LHS and HSS, may provide import clues on the physical driver for state transition, thus it is important to investigate the properties of the accretion flow during the source in the HIMS and SIMS.

In addition to the hard powerlaw and the thermal emission, the relativistic reflection spectrum is frequently reported \citep{fab1989, gar2014, pla2014, don2020a, don2020b, fen2022} in the X-ray spectrum of both black hole X-ray binaries (BHXRBs) and active galactic nucleus (AGNs). The reflection spectrum appears when a substantial flux of coronal photons are reflected from the surface of the disc. As a result, this reflected component includes absorption edges, fluorescent lines and a Compton hump. If the reflection emissions come from the region that is close enough to the black hole, it will be distorted by the relativistic effects, carrying the information of strong fields \citep{lao1991}. The study of reflection features can provide insights on the inclination, the iron abundance, and the ionization of the disc, as well as the geometry and the electron temperature of the corona. Moreover, the detailed modeling of reflection features is an important tool to measure the inner radius of the disc \citep{gar2015, xu2020, sri2020}. If the inner radius is located at the ISCO, the spin of the black hole can then be estimated \citep{bar1972}. The study of the reflection spectra in different states offers an opportunity to yield important insights on the co-evolution of the disc and corona. Interestingly, in contrast to the theoretical expectation \citep[e.g.,][]{esi1997, mey2000}, it has been suggested that the inner accretion disc is not truncated by moddelling the relativistic reflection in the LHS for some sources. For example, the inner radius is found to be very close to the ISCO for Cyg X-1 \citep{rei2010, par2015}, GX 339-4 \citep{gar2015, ste2017}, and MAXI J1820+070 \citep{bui2019}. Whether the truncation of the inner disc in LHS and at what phase the truncation appears are still in hot debate in recent years.

MAXI J1535-571 is an X-ray transient discovered in LHS by \emph{MAXI} \citep{neg2017a} and \emph{Swift} \citep{ken2017}, on September 2nd, 2017 (MJD 57999). Its X-ray spectral and timing properties \citep{neg2017b}, together with its bright radio signals \citep{res2017}, strongly suggest a black hole primary. MAXI J1535-571 was also observed in the optical and infrared bands \citep{sca2017, din2017}. Its X-ray spectra started to soften on September 10th \citep{nak2017, ken2017}, followed by the intermediate state which lasted for 2 months \citep{shi2017}. During the LHS-HIMS-SIMS transitions, low frequency QPOs were detected \citep{hua2018, ste2018, sti2018, sre2019}, and the evolution of compact jet and relativistic jet were reported \citep{rus2019, rus2020}. \citet{rus2019} also constrained the jet inclination to be $\leqslant45$ degrees. The source is heavily absorbed with a line-of-sight (LOS) column density larger than 10$^{22}$ ${\rm{cm}}^{-1}$ (\citealt{ste2018, cun2020}, and the references therein). The source distance ($D$) is estimated to be $4.1_{-0.5}^{+0.6}$ kpc based on the analysis on H\Romannum{1} absorption spectrum \citep{cha2019}. 

\citet{xu2018} analyzed the \emph{NuSTAR} data observed on September 7th during which MAXI J1535-571 was in the bright phase of the LHS. They found a strong relativistic reflection component in the \emph{NuSTAR} data. They reported no significant disc truncation and a rapidly rotating black hole (>0.84 and >0.987 with the {\tt relxilllpcp} and {\tt relxillcp} model, respectively). \citet{kon2020} found the spin was $0.7_{-0.3}^{+0.2}$ with the {\tt relxilllpcp} model using the data obtained by \emph{HXMT}, also on the September 7th, but the exposure time was less than 1 ks. \citet{mil2018} and \citet{sri2019} constrained the spin parameter using \emph{NICER} and \emph{AstroSat} observations, respectively. Both observations were obtained when the source was in the start of the HIMS. The best-fitting model of the \emph{NICER} data indicated a high spin of $0.994\pm0.002$, while \emph{AstroSat} data indicated a moderate spin of $0.67_{-0.04}^{+0.16}$. The inner radius and the spin are degenerate since they both affect the red wing of the fluorescent iron line. So, moderate spin may indicate a moderately truncated disc. The inconsistent values of spin also may attribute to the model difference.

Here we report a joint analysis of the \emph{NuSTAR} \citep[The Nuclear Spectroscopic Telescope Array,][]{har2013} and \emph{HXMT} (Hard X-ray Modulation Telescope, or \emph{Insight-HXMT}, \citealt{zha2014}) observations of MAXI J1535-571. We analyze 3 epochs data obtained as the source increased in intensity while undergoing transition from a bright-hard towards the soft state during the 2017 outburst. \emph{NuSTAR} is the first focusing high-energy X-ray telescope in orbit which covers a broad energy band (3-79 keV) with unprecedented energy resolution and sensitivity in the hard X-ray band. \emph{HXMT}, as the first Chinese X-ray astronomical satellite, includes three slat-collimated instruments: the Low Energy X-ray Telescope (LE, 1-15 keV), the Medium Energy X-ray Telescope (ME, 5-30 keV),
and the High Energy X-ray Telescope (HE, 20-250 keV). The dataset from both satellites are not affected by photon pile-up effects. Similar spectra analysis of Cyg X-1 \citep{zha2020} and MAXI J1820-070 \citep{you2021, zha2021, gua2021} using \emph{HXMT} have been reported. The paper is organized as follows: in Section \ref{sec:obs}, we describe the detail of observations and data reduction; the data analysis and results are presented in Section \ref{sec:ana}; Section \ref{sec:dis} includes discussions and Section \ref{sec:ccl} includes conclusions.

\section{Observations and Data reduction}
\label{sec:obs}
\begin{figure}
	\includegraphics[width=\columnwidth]{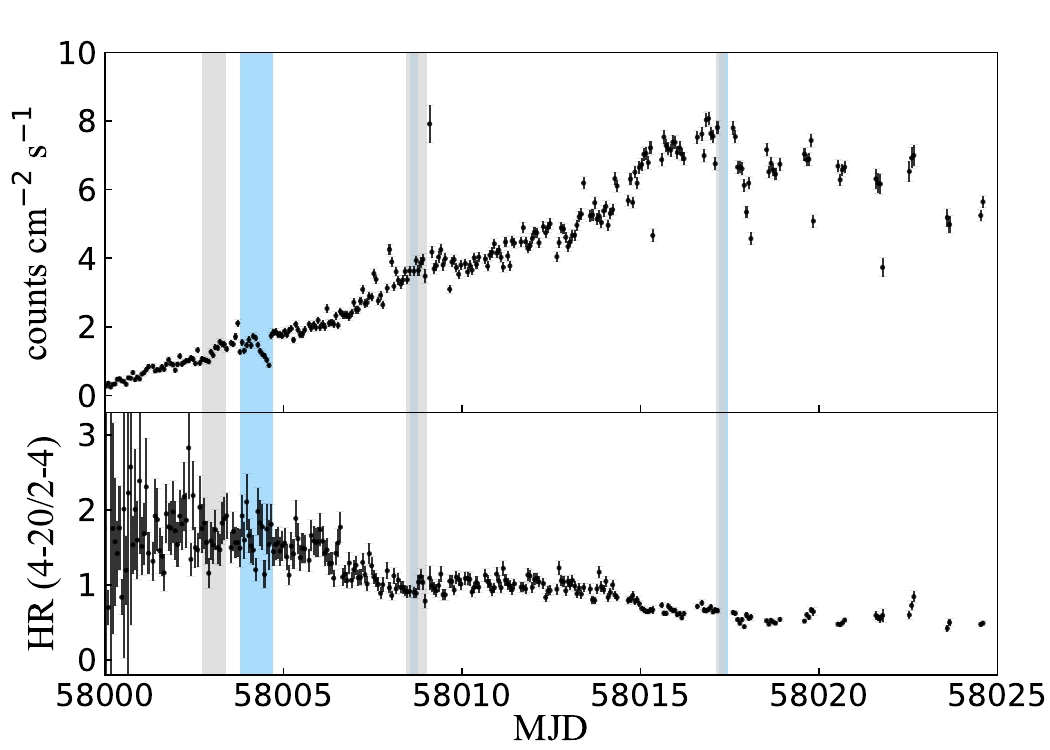}
    \caption{The upper panel shows the \emph{MAXI}/GSC 2-20 keV orbital light curve of the black hole binary MAXI J1535-571. The lower panel shows the \emph{MAXI} hardness ratio which is calculated as the ratio of the counts between 4-20 keV and 2-4 keV. The blue and gray shaded areas mark the \emph{NuSTAR} and \emph{HXMT} observations analyzed in this work.}
    \label{fig:lc}
\end{figure}

We show MAXI J1535-571's light curve and hardness ratio obtained by \emph{MAXI} \citep{mat2009} in Figure \ref{fig:lc}. In this paper, we analyze 3 \emph{NuSTAR} observations (blue shadow in Figure \ref{fig:lc}) with exposure time of 8685, 2258 and 1531 seconds, respectively. Quasi-simultaneous \emph{HXMT} observations are also analyzed with exposure time of 5614.5, 4316 and 3496 seconds, respectively. We mark \emph{HXMT} observations as gray shadow in Figure \ref{fig:lc}. For the data in Epoch 1 (ObsID: 90301013002), the \emph{NuSTAR} and \emph{HXMT} are not observed simultaneous with the \emph{NuSTAR} observation carried out less than 10 hours later than the \emph{HXMT} observation. However, the \emph{NuSTAR} and \emph{HXMT} data show roughly the same reflection spectral features (iron emission line and Compton hump region), suggesting no significant change in the reflection component, we thus still jointly analyze the \emph{NuSTAR} and \emph{HXMT} data to increase the signal-to-noise ratio (SNR). The including of the \emph{HXMT}/LE data enable us to perform spectral analysis down to 2.1 keV which will be valuable to detect the weak thermal emission in LHS. The details of observations of both satellites can be found in Table \ref{tab:obs}. The three epoch observations are in LH, HIMS and SIMS, respectively \citep{hua2018, tao2018}. In the following sections, we describe the observations and data reduction for \emph{NuSTAR} and \emph{HXMT}.

\begin{table*}
    \newcommand{\tabincell}[2]{\begin{tabular}{@{}#1@{}}#2\end{tabular}}
    \begin{threeparttable}[b]
	\centering
	    \begin{center}
	    \caption{Details of {\emph{NuSTAR}} and {\emph{HXMT}} observations}
    \label{tab:obs}
        \footnotesize
            \begin{tabular}{cccccccccc}
            \toprule
            \tabincell{c}{Mission}&Instrument&ObsID&MJD&Start Time&End time&\tabincell{c}{Exposure\\(s)}&\tabincell{c}{Count Rate$^a$\\ (cts s$^{-1}$)}&Total counts$^b$& State$^c$ \\            
        \midrule
\multicolumn{10}{c}{Epoch 1}\\
\emph{NuSTAR} &FPMA &90301013002$^d$  &58003.79  &09-07 18:41:09  &09-08 17:01:09   &8685   &$648.3 
  \pm 0.3$&1.11$\times10^7$&LHS\\
              &FPMB &... &... &...   &...   &9077   &$601.2  \pm 0.3$&&\\
\emph{HXMT}&LE& 11453500104  & 58002.72& 09-06 17:11:13  &09-06 20:22:09   &1047   &$269.5  \pm 0.5$&8.13$\times10^6$&\\ 
                   &ME& ... &... &...   &...   &1331   &$266.8  \pm 0.5$&&\\  
                   &HE& ... &... &...   &...   &1309   &$645.2  \pm 1.6$&&\\  
                   &LE&11453500105  &58002.85 & 09-06 20:22:09  &09-06 23:33:06   &	898	&$	280.0	\pm	0.6$	&&\\ 
                   &ME& ... &... &...   &...   &	1218	&$	274.0	\pm	0.5	$&&\\   
                   &HE& ... &... &...   &...   &1730	&$	652.6	\pm	1.3	$&&\\   
                   &LE&11453500106  &58002.98 & 09-06 23:33:06  &09-07 02:43:10   &1057	&$	289.6	\pm	0.5	$&&\\ 
                   &ME& ... &... &...   &...   &1193	&$	283.2	\pm	0.5	$&&\\   
                   &HE& ... &... &...   &...   &1543	&$	664.6	\pm	1.4	$&&\\
                   &LE&11453500107$^d$  &58003.11 & 09-07 02:43:10  &09-07 05:54:06   &937	&$	296.4	\pm	0.6	$&&\\
                   &ME& ... &... &...   &...   &1487	&$	287.5	\pm	0.5	$&&\\   
                   &HE& ... &... &...   &...   &1837	&$	666.2	\pm	1.2	$&&\\                   
                   &LE&11453500108  &58003.25 & 09-07 05:54:06  &09-07 09:05:03   &1676	&$	300.9	\pm	0.4	$&&\\
                   &ME& ... &... &...   &...   &1960	&$	293.6	\pm	0.4	$&&\\   
                   &HE& ... &... &...   &...   &378	&$	671.9	\pm	3.0	$&&\\                   
\midrule
\multicolumn{10}{c}{Epoch 2}\\  
\emph{NuSTAR} &FPMA &80302309002  &58008.55  &09-12 13:01:09  &09-12 18:26:09 &2258   &$1132.0 \pm 0.7$&5.05$\times10^6$&HIMS\\
              &FPMB &... &... &... &... & 2418   &$1032.0 \pm 0.7$&&\\
\emph{HXMT}&LE   &11453500144  &58008.44&09-12 10:38:15  &09-12 13:58:12   &1137	&$	979.3	\pm	0.9	$&3.73$\times10^3$&\\
                   &ME   &... &... &... &... &2168	&$	407.6	\pm	0.5	$&&\\    
                   &HE   &... &... &... &... &1684	&$	462.3	\pm	1.3	$&&\\    
                   &LE   &11453500145  &58008.58&09-12 13:58:12  &09-13 00:41:28   &3179	&$	996.2	\pm	0.6	$&&\\ 
                   &ME   &... &... &... &... &8202	&$	414.2	\pm	0.2	$&&\\ 
                   &HE   &... &... &... &... &9329	&$	474.1	\pm	0.6	$&&\\ 
\midrule
\multicolumn{10}{c}{Epoch 3}\\
\emph{NuSTAR} &FPMA &80302309010  &58017.21  &09-21 04:51:09  &09-21 10:46:09   &1531   &$1818.0 \pm 1.1$&5.50$\times10^6$&SIMS\\
              &FPMB &... &... &... &... &1652   &$1643.0 \pm 1.0$&&\\
\emph{HXMT} &LE &11453500901  &58017.10&09-21 02:26:26  &09-21 06:00:41   &1676	&$	2251.0	\pm	1.2	$&5.93$\times10^3$&\\  
                    &ME &... &... &... &... &2967	&$	399.3	\pm	0.4	$&&\\  
                    &HE &... &... &... &... &2467	&$	326.3	\pm	1.2	$&&\\  
                    &LE &11453500902  &58017.25&09-21 06:00:41  &09-21 09:21:07   &1820	&$	2259.0	\pm	1.1	$&&\\ 
                    &ME &... &... &... &... &2780	&$	387.4	\pm	0.4	$&&\\ 
                    &HE &... &... &... &... &3697	&$	326.3	\pm	0.9	$&&\\ 
            \bottomrule
            \end{tabular}
                \begin{tablenotes}
                \item \textbf{Note.} a. The count rate is shown within the energy band of 4.0-79.0 keV for  \emph{NuSTAR}/FPMA and FPMB, 2.1-10.0, 10.0-27.0, 27.0-60.0 keV for \emph{HXMT}/LE, ME, and HE, respectively. 
                
                b.The total number of counts are shown for the two modules of \emph{NuSTAR} and the three modules of \emph{HXMT} combined spectra.
                
                c. Spectral states labeled according to \citet{hua2018} and \citet{tao2018}.
                
                d. The \emph{NuSTAR} and \emph{HXMT} observations were analyzed by \citet{xu2018} and \citet{kon2020}, respectively.
                \end{tablenotes}
        \end{center}
    \end{threeparttable}
\end{table*}
\subsection{\emph{NuSTAR} Data Reduction}
\label{subsec:nu}
The \emph{NuSTAR} data were processed using \emph{NuSTAR} Data Analysis Software (NuSTARDAS v2.0.0) with \uppercase{CALDB} v20210524, which are included in \uppercase{HEASOFT} v6.28. We created cleaned event files using the \uppercase{NUPIPELINE} routine. The count rate exceeds 100 counts s$^{-1}$ in these 3 observations. Therefore, we set STATUEXPR to be ``STATUS==b0000xxx00xxxx000''. Especially for Obs. 1, we also set saacalc = 2, saamode = strict, and tentacle =NO to remove background flares, which is caused by enhanced solar activity. The X-ray spectra, backgrounds and instrument responses were generated using \uppercase{NUPRODUCTS}.
The spectra were extracted from a circular region with radius of 180$''$ centred on MAXI J1535-571, while backgrounds were extracted from a circular region with radius of 180$''$ located on the same detector. The spectra were grouped with \uppercase{grppha} to have at least 30 counts within an energy bin. We choose the 4-79 keV range for the spectral analysis.

\subsection{\emph{HXMT} Data Reduction}
\label{subsec:hxmt}
\emph{HXMT} includes the Low Energy X-ray Telescope (LE) , the Medium Energy X-ray Telescope (ME), and the High Energy X-ray Telescope (HE). We carried data reduction following the standard procedures for individual instruments, as the suggestions given by \emph{HXMT} team. The data pipelines and tools of \emph{HXMT} Data Analysis Software (HXMTDAS) v2.04 \footnote{http://hxmten.ihep.ac.cn/SoftDoc.jhtml} were used. The \emph{HXMT} spectra, were extracted based on the cleaned events files, which were filtered by the good time intervals (GTIs). The GTIs recommended by pipeline are intervals when (1) elevation angle greater than 10 degrees; (2) geomagnetic cut-off rigidities greater than 8 GeV; (3) satellite not in SAA and 300 seconds intervals near SAA; (4) pointing deviation to the source less than 0.04 degrees. We binned the spectra at least 30 counts within an energy bin. Then, the systematic uncertainties of 0.5\%/0.5\%/3\% were added for LE/ME/HE to account for the instrumental uncertainties{\footnote{The systematic uncertainties are related to the spectrum energy, and 1\%/2\%/3\% are recommended for LE/ME/HE. But we found that are overestimated for LE and ME (with a $\chi^2_{\nu}$ less than 1) when we compare the fit statistics between HXMT and NuSTAR. Therefore, the systematic error is set to 0.5\% for LE and ME.}}. For spectral analysis, we use 2.1-10 keV, 10-27 keV, 27-60 keV energy band for LE, ME and HE, respectively. The spectra at higher energies are dominated by the background.

Data in Epoch 1 was split into 5 continuous observations, and both Epoch 2 and 3 were split into 2. We used the following ways to check the spectral variability for all observations within one epoch. For each epoch, we performed a joint-fit with an absorbed powerlaw model to all spectra within this epoch. A constant multiplication factor was also included to account for the flux fluctuation. The parameters column density, photon index and normalization were linked among spectra. We find their data to model ratios are highly consistent. Additionally, if the photon index was allowed to be float among different observations, the value of it is consistent within the errors. Therefore, the source spectral shape is not significantly variable within each epoch. The average spectra were created with the \uppercase{addascaspec} tool for each of the 3 epochs, and were used for the subsequent spectral analysis.

\section{Spectral analysis and results}
\label{sec:ana}

The spectral analysis are performed using XSPEC version 12.11.1 \citep{arn1996}, which is included as part of the HEASOFT v6.28. In all models, a multiplicative constant model is included using the {\tt constant} model to account for the differences in the flux calibration between instruments. This constant is fixed at 1.0 for \emph{NuSTAR}/FPMA, and allowed to vary for \emph{NuSTAR}/FPMB and \emph{HXMT}/LE, ME, and HE, unless otherwise noted. We use the {\tt tbabs} model \citep{wil2000} to model the neutral Galactic absorption, with the abundances of \citet{wil2000} and the cross-sections of \citet{ver1996} adopted. All parameter uncertainties are quoted at 90\% confidence level for one parameter of interest. 

\subsection{Fitting spectra individually}
\label{subsec:FI}
\begin{figure*}
	\includegraphics[width=\textwidth]{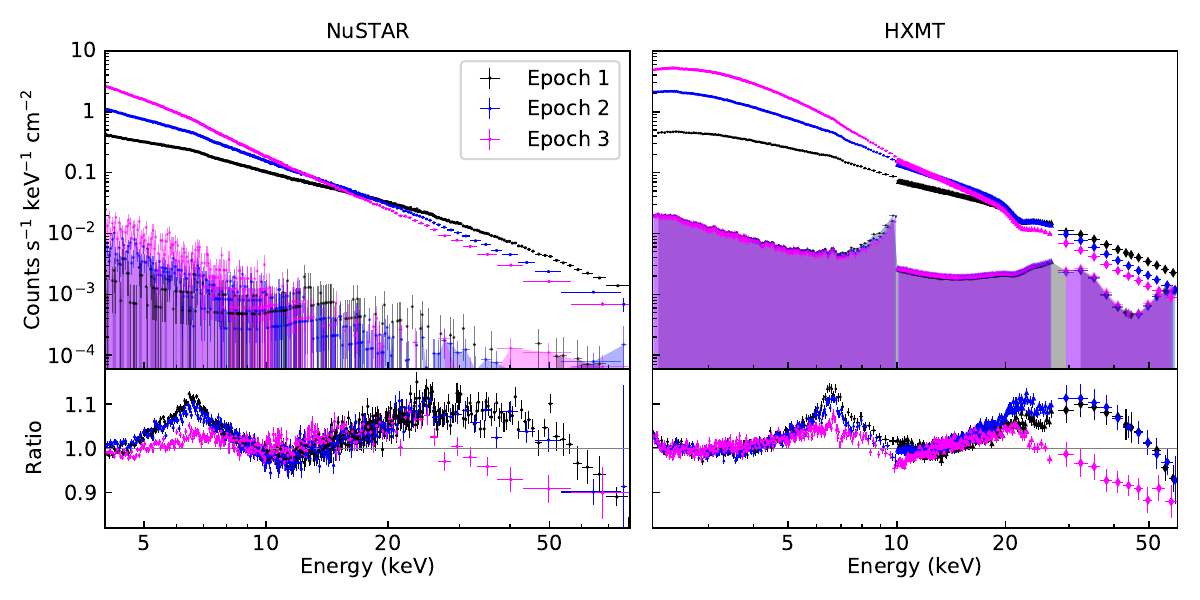}
    \caption{Spectra and ratios of the data-to-model consisting of a mult-temperature disc blackbody plus powerlaw. The \emph{NuSTAR} and \emph{HXMT} spectra were fitted together, but are shown in the left and right panels, respectively, for clarity. Background spectra using shaded regions are also shown in the upper panels. The spectra were fitted with the energy region 4-8 keV and 15-45 keV ignored. In all panels, Epoch 1 is shown in black, Epoch 2 is shown in blue, and Epoch 3 is shown in magenta. In the right panels, LE, ME and HE (3 instruments of \emph{HXMT}) data are shown in points, triangles and diamonds. The spectra have been rebinned to higher signal-to-noise in XSPEC for plotting purpose only. The residual structures indicate prominent reflection as seen by the broadened $\sim$6.5 keV Fe line and $\sim$20 keV Compton hump.}
    \label{fig:pl}
\end{figure*}
\begin{figure}
	\includegraphics[width=0.5\textwidth]{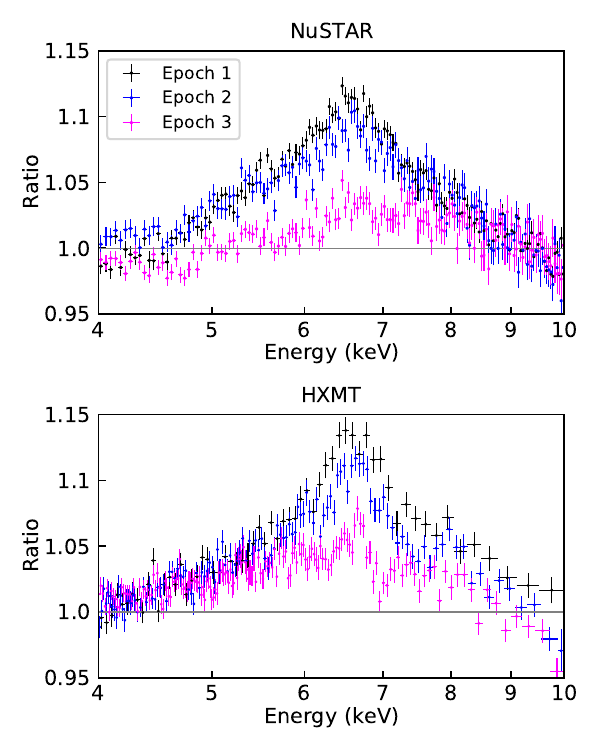}
    \caption{Close-up of iron line profiles in Figure \ref{fig:pl} for the \emph{NuSTAR} and \emph{HXMT} observations. In both panels, Epoch 1 is shown in black, Epoch 2 is shown in blue, and Epoch 3 is shown in magenta. The red wing appears to be relatively stable below $\sim$5 keV during 3 epochs.}
    \label{fig:Fe}
\end{figure}

We initially jointly fitted the \emph{NuSTAR} and \emph{HXMT} data for each of the 3 epochs with an absorbed powerlaw model plus a multi-temperature thermal disc ({\tt diskbb}) model \citep{mit1984}. There is no overlap in the coverage among the 3 instruments of \emph{HXMT}. We therefore linked the 3 constant parameters together to prevent the degeneracy between the constant parameters for LE/ME/HE and the photon index $\Gamma$ of powerlaw. The remained parameters were linked for the \emph{NuSTAR} and \emph{HXMT} data. The iron line region between 4-8 keV and Compton hump region between 15-45 keV were ignored to avoid potential contribution to the powerlaw continuum. We note that the residual profiles in the high energy band for \emph{NuSTAR} and \emph{HXMT} are slightly different for Epoch 1. We find that this could be due to the change of the photon index $\Gamma$ of the the powerlaw continuum component between the \emph{NuSTAR} and \emph{HXMT} observations in Epoch 1. Indeed, the residuals for the \emph{NuSTAR} and \emph{HXMT} data are consistent if the photon index $\Gamma$ is fitted independently for the \emph{NuSTAR} and \emph{HXMT} data. Figure \ref{fig:pl} shows the ratios of the data-to-model for 3 epochs. It is clear that significant reflection features are revealed in all of 3 epochs. The profile of the iron line appears to be relatively stable with its red wing extending below $\sim$5 keV over 3 epochs, indicating that the inner accretion disc may always be at the ISCO. On the other hand, the flux of the iron line decreases from Epoch 1 to 3 (Figure \ref{fig:Fe}).

We then replaced the powerlaw model with a reflection model {\tt relxillCp} (relxill v1.4.3, \citealt{dau2014, gar2014}) to fit the relativistically blurred reflection component in the data. The {\tt relxillCp} model also internally includes a continuum component which is calculated using the Comptonization model {\tt nthcomp} \citep{zdz1996, zyc1999}. A distance reflection component, which is generally believed to originate from the reflection of the outer accretion disc, is also added using the model {\tt xillverCp} \citep{gar2010}. The total model is given by {\tt constant*tbabs(diskbb+relxillCp+xillverCp)} in XSPEC. We fitted each epoch independently with this model.

In order to test the potential evolution of the inner radius of the disc over the rise phase of the outburst, we fixed the spin ($a_*$) of the black hole at its maximal value 0.998 ($R_{\rm{ISCO}} = 1.235 R_{\rm{g}}$). While the inner radius ($R_{\rm{in}}$) of the disc was free in the {\tt relxillCp} model. The outer radius of the disc ($R_{\rm{out}}$) was fixed at the default value 400 $R_{\rm{g}}$. We found that the best-fitting value of the iron abundance parameter $A_{\rm{Fe}}$ was very close to the solar abundance in all the 3 epochs. We thus fixed $A_{\rm{Fe}}$ at the solar abundance. The emissivity profile is described by a broken powerlaw in {\tt relxillCp}, i.e., $\epsilon(r)\propto r^{-q_\text{in}}$ for $r<R_{\text{br}}$ and $\propto r^{-q_\text{out}}$ for $r>R_{\text{br}}$. $R_{\text{br}}$ is the break radius, while $q_{\rm{in}}$ and $q_{\rm{out}}$ are the index for the inner and outer regions, respectively. Our data, however, cannot constrain all the 3 parameters simultaneously. We thus used a simple powerlaw to describe the emissivity profile by linking the value of $q_\text{out}$ to that of $q_\text{in}$. All the other parameters (the inclination angle $i$, the ionization state log${\rm{xi}}$, the electron temperature $kT_{\rm{e}}$, and the normalization $N_{\rm{rel}}$) in the {\tt relxillCp} are free parameters.

The parameters in the {\tt xillverCp} component were linked to those of {\tt relxillCp} except for the ionization ($\log\xi_{\rm{xil}}$) and the normalization ($N_{\rm{xil}}$) paramaters. We initially fixed the $\log\xi_{\rm{xil}}$ of the {\tt xillverCp} component at zero.  Compared with the best-fitting results shown in Table \ref{tab:FI}, the statistics were degraded by $\Delta \chi^2 = 84.5$, 93.7, and 52.5 for 1 degree of freedom in Epoch 1, 2, and 3, respectively. The case that the distant reflection comes from ionized material is consistent with the previous studies \citep{xu2018, sri2019}, in which they reported that the ionization of the outer disc could be high (e.g. $\log\xi_{\rm{xil}}>>0$) and may be different from the inner disc region. We then set $\log\xi_{\rm{xil}}$ of the {\tt xillverCp} component independent of that ($\log\xi_{\rm{rel}}$) of {\tt relxillCp}. The two values of $\log\xi$ are consistent within uncertainty for the Epoch 3, but they are quite distinct for the Epoch 1 and 2. Therefore, the two values of $\log\xi$ in the {\tt xillverCp} and {\tt relxillCp} were linked for Epoch 3, while they were fitted independently for Epoch 1 and 2. In addition, we find that unlinking photon index ($\Gamma$) of {\tt powerlaw} between the \emph{NuSTAR} and the \emph{HXMT} data can greatly improve the fitting result for Epoch 1 with $\Delta \chi^2 = 387.52$ for one additional parameter. The values of photon index are constrained to be $1.93_{-0.02}^{+0.01}$ and $1.89_{-0.02}^{+0.01}$, respectively. The minor difference between values of $\Gamma$ maybe attributed to the non-strictly simultaneous observation in Epoch 1.

This model can fit all the 3 epochs well, yielding reasonable statistics with $\chi^2/\nu$ = 4576.52/4273 = 1.07, 3296.6/3323 = 0.99 and 3317.74/2991 = 1.11 for the Epoch 1, 2 and 3, respectively. Table \ref{tab:FI} presents the details of independent fitting of each epoch. The best-fitting emissivity profile is steep in each of the 3 epochs, i.e. $q > 9.81$ for Epoch 1, $q = 7.33_{-0.97}^{+1.44}$ for Epoch 2, and $q > 7.22$ for Epoch 3. \citet{xu2018} also used the same model to fit the \emph{NuSTAR} data of Epoch 1. Our best-fitting results of Epoch 1 are consistent with their results, albeit they fixed $q_{\rm{out}}$ at 3 with $R_{\rm{br}}$ = $10R_{\rm{g}}$. Assuming the Newtonian case \citep{ss1973, nov1973, rey1997}, i.e. $q$ was fixed at 3, we got much worse fits with $\Delta\chi^2$ = 97.67, 52.23, and 38,87 for Epoch 1, 2 and 3, respectively. 
 
The inferred LOS hydrogen column density ($N_{\rm{H}}$) varies from  $\sim6.95\times 10^{22}$ ${\rm{cm}}^{-1}$ for Epoch 1 to $\sim5.4\times 10^{22}$ ${\rm{cm}}^{-1}$ for Epoch 2 and 3. These best-fitting LOS column densities are higher than that expected from Galactic absorption ($1.40 \times 10^{22}$ ${\rm{cm}}^{-1}$), which may imply intrinsic absorption from the source. The variation of this excess absorption could attribute to the wind or the outer region of the disc. However, it can also be due to systematic uncertainty with the model, as such variation of the absorption is not typical in low-mass X-ray binaries. In addition, there is no evidence of the absorption lines detected in the X-ray spectra of MAXI J1535-571. To test whether a constant LOS absorption can fit the data, we performed a simultaneous fit to all the 3 epochs. 

\begin{figure*}
	\includegraphics[width=\textwidth]{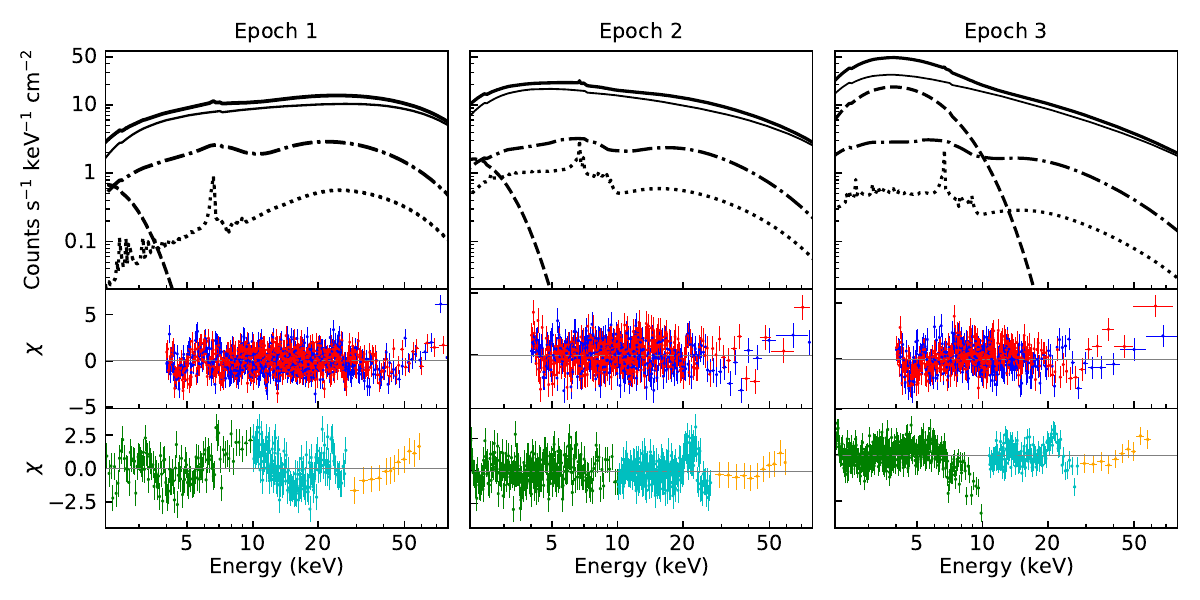}
    \caption{Top panels: the best-fitting models for MAXI J1535-571. We fit 3 epochs simultaneously, but show the results for each of the epoch in an individual way for clarity. The total model is shown in thick-solid line; the thermal emission ({\tt diskbb}) from the disc is shown in dashed line; the Compotonization component {\tt nthcomp} is shown in dot-dashed line, and it is calculated internally by {\tt relxillcp}; the relativistic and distant reflection are shown in dotted and thin-solid lines. For Epoch 1, the models for \emph{HXMT} is very similar to that for \emph{NuSTAR}, so only the lines for \emph{NuSTAR} are presented for visual clarity. Middle and bottom panels: \emph{NuSTAR} and \emph{HXMT} 
   residuals with 1$\sigma$, respectively. The blue and red are for FPMA and FPMB. The green, cyan, and orange are for LE, ME, and HE, respectively. The data has been rebinned here for display clearly.}
    \label{fig:rel}
\end{figure*}
\renewcommand{\arraystretch}{1.2}
\begin{table*}
\begin{threeparttable}[b]
\begin{center}
\caption{Best-fitting Parameters to \emph{NuSTAR} and \emph{HXMT} Spectra}
\begin{tabular}{ccccccccc}
\toprule
Component&Parameter&\multicolumn{3}{c}{M1}&&\multicolumn{3}{c}{M2}\\
\cline{3-5}
\cline{7-9}
&&Epoch 1 &Epoch 2&Epoch 3&&Epoch 1&Epoch 2&Epoch 3\\
&&(LHS)&(HIMS) &(SIMS)&&(LHS) &(HIMS) &(SIMS)\\
\midrule
{\tt TBabs}&$N_{\rm{H}}$ ($\times10^{22}$ cm$^{-2}$)& \multicolumn{3}{c}{5.48$_{-0.04}^{+0.05}$}&&\multicolumn{3}{c}{5.46$_{-0.02}^{+0.03}$}\\
{\tt diskbb}&$T_{\rm{in}}$ (keV)&0.324$\pm$0.012&0.349$_{-0.015}^{+0.013}$&1.2$_{-0.008}^{+0.004}$&&0.326$_{-0.008}^{+0.003}$&0.34$_{-0.003}^{+0.006}$&1.201$_{-0.002}^{+0.001}$\\
&$N_{\rm{disc}}$ ($\times 10^{3}$)&167.49$_{-42.46}^{+59.08}$&225.27$_{-50.41}^{+79.51}$&1.91$_{-0.03}^{+0.04}$&&54.71$_{-13.51}^{+29.52}$&265.76$_{-19.21}^{+44.03}$&1.92$_{-0.02}^{+0.01}$\\
{\tt relxillCp}&$q$&$>$9.13&8.73$_{-0.67}^{+0.57}$&$>$9.5&&7.49$_{-0.41}^{+0.33}$&8.68$_{-0.48}^{+0.38}$&$>$9.72\\
&$a_*$&\multicolumn{3}{c}{...}&&\multicolumn{3}{c}{0.985$_{-0.004}^{+0.002}$}\\
&$i$ (degrees)&\multicolumn{3}{c}{72.8$_{-1.28}^{+0.86}$}&&\multicolumn{3}{c}{70.7$_{-0.48}^{+0.17}$}\\
& $R_{\rm{in}}$ ($R_{\rm{g}}$)&$1.51_{-0.03}^{+0.04}$&$1.38_{-0.03}^{+0.02}$&1.38$_{-0.02}^{+0.03}$&&\multicolumn{3}{c}{...}\\
& $\Gamma$&1.822$_{-0.023}^{+0.011}$&2.401$_{-0.01}^{+0.012}$&2.795$_{-0.007}^{+0.006}$&&1.819$_{-0.003}^{+0.002}$&2.401$\pm$0.002&2.79$_{-0.001}^{+0.002}$\\
& log${\rm{\xi}_{\rm{rel}}}$ (erg cm s$^{-1}$)&3.24$_{-0.08}^{+0.27}$&3.11$\pm$0.06&3.3$_{-0.05}^{+0.07}$&&3.25$_{-0.02}^{+0.01}$&3.13$_{-0.01}^{+0.03}$&3.3$_{-0.03}^{+0.02}$\\
& $kT_{\rm{e}}$ (keV)&18.13$_{-0.29}^{+0.35}$&34.01$_{-2.08}^{+1.98}$&$>$322.3&&18.06$\pm$0.15&34.09$_{-1.04}^{+1.07}$&$>$311.01\\
&$R_{\rm{f}}$&0.9$_{-0.07}^{+0.06}$&1.08$_{-0.13}^{+0.12}$&0.95$_{-0.09}^{+0.16}$&&0.79$_{-0.04}^{+0.03}$&0.87$\pm$0.06&0.7$_{-0.05}^{+0.02}$\\
&$N_{\rm{rel}}$ ($\times 10^{-2}$)&9.62$_{-1.36}^{+0.2}$&39.88$_{-0.85}^{+1.05}$&134.35$_{-6.28}^{+1.96}$&&9.52$_{-0.13}^{+0.02}$&40.68$_{-0.17}^{+0.08}$&134.54$_{-0.46}^{+0.94}$\\
& $\Gamma_{\rm{HXMT}}$&1.784$_{-0.02}^{+0.01}$&...&...&&1.78$\pm$0.004&...&...\\
{\tt xillverCp}& log${\rm{\xi}}_{\rm{xil}}$ (erg cm s$^{-1}$)&2.53$_{-0.13}^{+0.14}$&$3.66_{-0.15}^{+0.14}$&...&&2.8$\pm$0.04&3.58$_{-0.15}^{+0.08}$&...\\
& $N_{\rm{xil}}$ ($\times 10^{-2}$)&1.54$_{-0.19}^{+0.21}$&6.58$_{-1.04}^{+1.7}$&16.0$_{-3.82}^{+4.27}$&&1.94$_{-0.12}^{+0.09}$&6.23$_{-0.35}^{+0.6}$&17.47$_{-3.26}^{+2.29}$\\
{\tt constant}&$C_{\rm{FPMB}}$&1.022$\pm$0.001&1.007$\pm$0.001&0.997$\pm$0.001&&1.022$\pm$0.001&1.007$_{-0.001}^{+0.002}$&0.997$\pm$0.001\\
&$C_{\rm{LE}}$&0.745$\pm$0.003&0.955$\pm$0.002&0.976$\pm$0.001&&0.746$\pm$0.002&0.955$_{-0.001}^{+0.002}$&0.975$\pm$0.001\\
&$C_{\rm{ME}}$&0.741$\pm$0.003&0.96$\pm$0.002&0.939$\pm$0.002&&0.741$\pm$0.002&0.96$\pm$0.002&0.94$\pm$0.002\\
&$C_{\rm{HE}}$&0.795$\pm$0.012&0.972$\pm$0.015&0.955$\pm$0.015&&0.795$_{-0.011}^{+0.012}$&0.968$_{-0.014}^{+0.015}$&0.947$\pm$0.015\\
\midrule
&$\chi^2$&\multicolumn{3}{c}{11296.6}&&\multicolumn{3}{c}{11356.31}\\
&$\nu$&\multicolumn{3}{c}{10591}&&\multicolumn{3}{c}{10593}\\
&$\chi^2_{\nu}$&\multicolumn{3}{c}{1.07}&&\multicolumn{3}{c}{1.07}\\
\bottomrule
        \end{tabular}
        \begin{tablenotes}
        \item \textbf{Notes.} The model {\tt constant*tbabs(diskbb+relxillCp+xillverCp)} is used to fit in M1 and M2. In M1: the spin parameter $a_*$ is fixed at 0.998, the inner radius $R_{\rm{in}}$ is free. In M2: $a_*$ is free and linked among 3 epochs, $R_{\rm{in}}$ is fixed at -1, which means $R_{\rm{in}}$ = $R_{\rm{ISCO}}$. In the two models, the column density $N_{\rm{H}}$ and inclination angle $i$ are linked among 3 epochs. The emissivity profile is assumed to be a single powerlaw, for which the emissivity index $q_{\rm{in}}$ = $q_{\rm{out}}$ = $q$. The constant factor is fixed at unity for \emph{NuSTAR}/FPMA, and free for \emph{NuSTAR}/FPMB ($C_{\rm{FPMB}}$), \emph{HXMT}/LE ($C_{\rm{FPMB}}$), ME ($C_{\rm{FPMB}}$), and HE ($C_{\rm{FPMB}}$). The other free parameters listed above: Temperature of the disc ($T_{\rm{in}}$); Inner radius of the disc ($R_{\rm{in}}$); Photon index ($\Gamma$ and $\Gamma_{\rm{HXMT}}$, which is only different for Epoch 1); Ionization state  ($\mathrm{log}\xi_{\rm{rel}}$ and $\mathrm{log}\xi_{\rm{xil}}$ for {\tt relxillcp} and {\tt xillvercp}, respectively, which is linked for Epoch 3); Electron temperature ($kT_{\rm{e}}$); Reflection fraction ($R_{\rm{f}}$); Normalization constants of {\tt diskbb} ($N_{\rm{disc}}$), {\tt relxillcp} ($N_{\rm{rel}}$) and {\tt xillvercp} ($N_{\rm{xil}}$).
        \end{tablenotes}
\label{tab:JF}
\end{center}
\end{threeparttable}
\end{table*}

\subsection{Fitting spectra simultaneously}
\label{subsec:JF}
The LOS column density $N_{\rm{H}}$, inclination angle of the disc $i$, Fe abundance $A_{\rm{Fe}}$, and the black hole spin $a_*$ were not supposed to vary among the 3 epochs, they were thus linked together. We again fixed $A_{\rm{Fe}}$ at solar abundance and $a_*$ at the maximum 0.998. The $i$ was left as free parameter. The remaining parameters were fitted independently for each epoch, and were set up as stated in Section \ref{subsec:FI}. We refer this model as M1. The model can fit the data well with $\chi^2/\nu$ = 11296.6/10591 = 1.07. The best-fitting parameters for M1 are presented in Table \ref{tab:JF}.
The components of the model together with the residuals are shown in Figure \ref{fig:rel}. A positive feature between 21-23 keV in \emph{HXMT} residuals (bottom panels of Figure~\ref{fig:rel}), which is a known effect that is caused by the photoelectric effect of silver elements \citep{you2021}. Ignoring this energy range (only for the \emph{HXMT}/ME instrument) does not affect the results. We also note that an excess at high energy tail is shown in the residuals in Epoch 1. Such excess may be attribute to the weaker disc component in the joint-fit, as it is not seen when fit the spectra individually (Section \ref{subsec:FI}) of which we got a stronger disc component with also a slightly larger $kT_{\rm{e}}$ and higher column density. Since the column density is expected to not vary dramatically among these 3 epochs. We therefore mainly report on the results from the joint-fit to 3 epochs (M1).

The best-fitting $N_{\rm{H}}$ is found to be ($5.48_{-0.04}^{+0.05})\times 10^{22}$ ${\rm{cm}}^{-1}$. This high absorption is slightly larger than the result obtained from the \emph{Swift} ($\sim3.6\times 10^{22}$ ${\rm{cm}}^{-1}$, \citealt{ken2017}) and \emph{NICER} ($\sim4.05\times 10^{22}$ ${\rm{cm}}^{-1}$, \citealt{gen2017}; $\sim4.89\times 10^{22}$ ${\rm{cm}}^{-1}$ \citealt{mil2018}). It is noteworthy that fixing $N_{\rm{H}}$ at a smaller value will significantly worsen the fit, e.g., $\Delta\chi^2=186.91$ for 1 degree of freedom if $N_{\rm{H}}$ is fixed at $5\times 10^{22}$ ${\rm{cm}}^{-1}$.
    
The best-fitting photon index in Epoch 1 is $\Gamma = 1.822_{-0.023}^{+0.011}$ ($\Gamma = 1.784_{-0.020}^{+0.010}$ for \emph{HXMT} spectra). The best-fitting $\Gamma$ increases to $2.401_{-0.010}^{+0.012}$ and $2.795_{-0.007}^{+0.006}$ in Epoch 2 and 3, respectively. The temperature of the disc does not change significantly ($\sim0.3-0.4$ keV) in Epoch 1 and 2, while it becomes much higher ($\sim1.2$ keV) in Epoch 3. This trend is clearly illuminated in the top panel of Figure \ref{fig:rel} where we showed the best-fitting model for each component. The relatively weak thermal components showed in Epoch 1 and 2 are comparable, while a prominent thermal component is clearly shown in Epoch 3 (dashed line in the upper panels). We find that a steep emissivity index is required for all the 3 epochs ($q > 9.13$ for Epoch 1, $q = 8.73_{-0.67}^{+0.57}$ for Epoch 2, and $q > 9.50$ for Epoch 3). The parameter $kT_{\rm{e}}$, which represents the temperature of the electrons in the corona, is constrained to be $18.13_{-0.29}^{+0.53}$ keV for Epoch 1, $34.01_{-2.08}^{+1.98}$ keV for Epoch 2, and $ > 322.30$ keV for Epoch 3. The reflection fraction $R_{\rm{f}}$ is approximate unity in all 3 epochs, indicating that half of the powerlaw photons irradiate the disc.  The constant factor is low for LE/ME/HE of \emph{HXMT} in Epoch 1, which is because of non-simultaneity. In addition, the inner radius of the accretion disc is broadly consistent among the 3 epochs with only minor difference, i.e., $R_{\rm{in}} = 1.51_{-0.03}^{+0.04}~R_{\rm{g}}$ for Epoch 1, $R_{\rm{in}} = 1.38_{-0.03}^{+0.02}~R_{\rm{g}}$ for Epoch 2, and $R_{\rm{in}} =1.38_{-0.02}^{+0.03}~R_{\rm{g}}$ for Epoch 3. However, the best-fitting inclination angle, $i$ = $72.80_{-1.28}^{+0.86}$ degrees, is much higher than that measured from radio jet. 

Our results suggests that the inner radius of the accretion disc does not change significantly which may indicate that it extends to the ISCO in all the three accretion states studied in this work. To self-consistently measure the spin of black hole, we then fixed the inner radius for all 3 epochs at ISCO ($R_{\rm{in}} = -1$) in the {\tt relxillcp} model. The $a_{*}$ was allowed to be free, but linked together among 3 epochs. We refer this model as M2. Comparing to M1, M2 can equally fit all the data well with $\chi^2/\nu = 11356.31/10593 = 1.07$. We present the best-fitting parameters in Table \ref{tab:JF}. We obtained a precise measurement of the spin ($a_{*}$ = $0.985_{-0.004}^{+0.002}$). The $R_{\rm{f}}$ is slightly lower than that in M1. The values of other parameters are similar to those obtained with M1. 

We also tried to fit the data with the lamp-post model {\tt relxilllpcp}. In this scenario, the hard X-ray photons are produced in a point source above the black hole spin axis \citep{min2004}. The height ($h$) of the corona, instead of $q_{\rm{in}}$, $q_{\rm{out}}$ and $R_{\rm{br}}$ in {\tt relxillcp}, is used to describe the illumination of the disc. The lamp-post configuration with a low height of the corona have been successfully used to explain the steep emissivity profile found in several BHXRBs \citep{dur2016, gar2018}. We jointly fit all the 3 epoch data together, allowing the height parameter ($h$) to be vary among the 3 epochs (the variant of M1). The other parameters set are as M1. This model provides a slightly worse fit to the data comparing with M1, with $\Delta \chi^2 = 163.55$ for the same degree of freedom. The $R_{\rm{in}}$ can not be well constrained in the lamp-post configuration. The obtained $R_{\rm{in}}$ for Epoch 1, 2 and 3 are < 4.61, < 13.94, and < 8.54 in units of $R_{\rm{g}}$, respectively. The $h$ are estimated to be $37.68_{-6.11}^{+7.14}$, $44.06_{-6.72}^{+12.38}$, and $20.33_{-1.88}^{+1.55}$ in units of $R_{\rm{g}}$ for 3 epochs, respectively, which is not consistent with a compact corona close to the black hole. The inclination angle is $\sim$66-70 degrees. 

The result presented in \citet{ste2017} emphasises the importance of the Compton scattering of the reflected photons by the hot coronal. To take this effect into account, we built a model in which the model {\tt simpl}, a kernel to calculate the Compton scattering \citep{ste2009}, is used to convolve the thermal and reflected photons. In {\tt simpl}, the free parameters are the scattered fraction $f_{\rm{sc}}$ and the $\Gamma$. The $f_{\rm{sc}}$ represents the proportion of the seed photons being scattered. However, we find that the $f_{\rm{sc}}$ is poorly constrained. It doesn't show any significant influence on the calculation of the relativistic reflection. So we opt for M1 instead.

\section{Discussions}
\label{sec:dis}
We have presented the detailed multi-epoch analysis of the reflection spectra of the black hole binary candidate MAXI J1535-571 over its rising phase of the outburst in 2017. The data were observed quasi-simultaneously by \emph{NuSTAR} and \emph{HXMT} when the source was in the LHS (Epoch 1), HIMS (Epoch 2), and SIMS (Epoch 3). We initially fitted the 3 epochs independently. We then performed joint modelling of the data for the 3 epochs. After subtracting the continuum (absorbed thermal emission plus power law component), prominent reflection features including the relativistic Fe K$\alpha$ line and the Compton hump are detected in each of the 3 epochs. The Fe K$\alpha$ line profile does not change significantly among the 3 epochs, while its flux decreases gradually from Epoch 1 to 3. In addition to the smeared reflection from the disc close to the black hole, the distant reflection, which maybe reflected from the ionized surface of the outer disc or the companion, was also observed. The {\tt relxillcp} and {\tt xillvercp} models were used in this work to fit the relativistic and distant reflection, respectively. We found that the hydrogen column density changes when fitted the 3 epochs independently, which may be induced by the systematic issues. Therefore, we also fitted the data from the 3 epochs simultaneously with the hydrogen column density assumed to be the same. 

The inclination of the accretion disc measured by modeling the X-ray reflection spectra is high ($\sim$70 degrees) in MAXI J1535-571. This is in agreement with the previous results by fitting \emph{NuSTAR}, \emph{NICER}, and \emph{AstroSat} data in \citet{xu2018}, \citet{mil2018}, and \citet{sri2019}, respectively. The inclination we obtained is significantly larger than the jet inclination ($\leqslant$ 45 degrees) which is measured by analyzing the radio data \citep{rus2019}. We also tried to fit the data with the inclination fixed at smaller values, e.g., less than 45 degrees. However, this always resulted in an unacceptable fit. Our results imply that the rotation axis of the inner accretion disc seems to be misaligned with the radio jet. Additionally, the jet and binary orbital plane is potentially misaligned. Such discrepancy has been previously reported in other systems like Cyg X-1 \citep{tom2014, par2015, wal2016} and MAXI J1820+070 \citep{pou2022}. 

The reflection-based measurements constrained the spin of the black hole in MAXI J1535-571 to be $>0.987$ \citep{xu2018}, $0.994 \pm 0.002$ \citep{mil2018}, $0.7_{-0.3}^{+0.2}$ \citep{kon2020}, and $0.67_{-0.04}^{+0.16}$ \citep{sri2019} by analysing data obtained between September 7 and 13. Because of the strong degeneracy between the spin and the inner radius, the intermediate spin measured in the HIMS by \citet{sri2019} may indicate that the disc is moderately truncated before it extending down to the ISCO. In this work, we use the \emph{NuSTAR} and \emph{HXMT} data observed on September 7 (LHS), 12 (HIMS) and 21 (SIMS) to study the potential evolution of the disc inner radius in MAXI J1535-571. We find that the inner radius does not change significantly in the three epochs with $R_{\rm{in}}\lesssim 1.55 R_{\rm{g}}$. The lack of the disc truncation is inconsistent with the work by \citet{sri2019}, which may attribute to the high iron abundance assumed in their model. The phenomenon of the disc inner radius without receding or proceeding represents $R_{\rm{in}} = R_{\rm{ISCO}}$. The spin is estimated to be $0.985_{-0.004}^{+0.002}$ via letting $a_{*}$ free instead of $R_{\rm{in}}$, suggesting a rapidly rotating black hole in MAXI J1535-571, which is in agreement with \citet{xu2018} and \citet{mil2018}. 

Except the two key systematic parameters, i.e., the spin of the black hole and the inclination of the inner disc, were estimated, the properties of the thermal emission and the Comptonized component are also explored. The parameters related to them present good consistency between M1 and M2. We note that these two components are exhibiting evolution. Epoch 1 is obtained when the source is in the bright phase of the hard state, while Epoch 2 is obtained at the beginning of the hard-to-soft state. In epoch 3, of which the luminosity of the source is approaching the peak during the outburst, and the source stay in the soft intermediate state. From Figure \ref{fig:rel}, the flux is dominated by the powerlaw component in Epoch 1 and 2, while the powerlaw and thermal components are equivalently strong in Epoch 3. 

The thermal emission observed above 2.1 keV is equally weak in Epoch 1 and 2, but becomes strong in Epoch 3. As listed in Table \ref{tab:JF}, the two best-fitting parameters, $T_{\rm{in}}$ and $N_{\rm{disc}}$ ($N_{\rm{disc}} = (r_{\rm{in}}/D)^2\times\cos i$)\footnote{To distinguish from the inner radius $R_{\rm{in}}$ shown in {\tt relxillcp} model, we use $r_{\rm{in}}$ to represent the inner radius indicated by {\tt diskbb}.}, of model {\tt diskbb} change significantly during the source transited from Epoch 2 to 3, which is in agreement with the results in \citet{tao2018}. It appears that the inner radius of the disc is slightly truncated in Epoch 1 and 2, which is inconsistent with the stable inner radius by modelling relativistic reflection components. Based on the results in Table \ref{tab:JF}, we calculated the unabsorbed disc flux (0.001-20 keV) for the 3 epochs, which are $\sim3.97$, $\sim7.22$, $\sim8.58$ (in units of $\times10^{-8}$ ergs cm$^{-2}$ s$^{-1}$), respectively. The flux does not show significant change from Epoch 2 to Epoch 3. This is inconsistent with the rise of the count rate shown in Figure \ref{fig:lc}, which indicates the increase of the accretion rate. The similar flux of the disc emission maybe led by the model without accounting for the Compotonization of the disc photons in corona. In the other hand, the effective temperature and the effective radius should be estimated after correcting the $T_{\rm{in}}$ and $r_{\rm{in}}$ by a hardening factor $f$. Because the model {\tt diskbb} does not account for any effects from the general relativity or electron scattering. A positive correlation between $f$ and accretion rate was reported in \citet{dav2019} and \citet{don2008}. Therefore, the abrupt change in the $T_{\rm{in}}$ and $N_{\rm{disc}}$ may be attributed from the change of accretion rate and hardening factor.

The photon index of $\Gamma\sim1.82$ in Epoch 1 is typical of the hard state. The spectrum softens as the state transition progresses, in Epoch 2 ($\Gamma\sim$2.40) and Epoch 3 ($\Gamma\sim$2.79). $\Gamma$ is used to describe the slope of the powerlaw, a component produced by inverse Compton scatter of the thermal emission in the corona. $\Gamma$ is related to the electron temperature ($kT{\rm{e}}$) and optical depth ($\tau$) of the corona by formula \citep{zdz2020}:
\begin{equation}
    \Gamma =-\frac{1}{2}+\sqrt{\frac{9}{4}+\frac{1}{u\theta(1+\theta+3\theta^2)}} 
    \label{equ:1}
\end{equation}
where $\theta$ is determined by $kT{\rm{e}}/m_ec^2$, and $m_ec^2$, the rest mass of the electron, is equal to 511keV. The term $u$ is the average number of scattering, which is calculated as follows \citep{zdz2020} :
\begin{equation}
    u = \tau(a+b\tau)
    \label{equ:2}
\end{equation}
\begin{equation}
    a = \frac{1.2}{1+ \theta + 5\theta^2}, b = \frac{0.25}{1+ \theta + 3\theta^2}
\end{equation}

The corona temperature, however, is challenging to be determined by X-ray spectral analysis because of the low sensitivity of the detectors at high energies (> 10 keV), until the launch of the \emph{NuSTAR} mission. Indeed, \emph{NuSTAR} observations have provided opportunities to detect the $kT{\rm{e}}$ in a large number of AGNs and X-ray binaries \citep{loh2015, pah2017, lan2019, yan2020}. On the other hand, a better statistics can be achieved by adding \emph{HXMT} observations in this work. The values of $kT{\rm{e}}$ are constrained to be $\sim18$, $\sim34$, and $> 311$ keV for Epoch 1, 2, and 3 respectively. The unconstrained upper limit in Epoch 3 maybe caused by the extremely steep of the powerlaw and the signal to noise is not sufficient enough. The corona temperature change slightly from Epoch 1 to 2, but increases more than ten times in Epoch 3.

The increased $\Gamma$ represents that the spectrum is becoming softer. If it is the case discovered in the LHS, the softening can be explained by the movement of the inner disc towards the black hole or the inflowing corona with a moderately relativistic velocity \citet{zdz1999}. More soft photons emitted from the disc will go into the Comptonization region. This will increase the cooling effect of the population of electrons, then the $kT{\rm{e}}$ of the corona decreases and the $\Gamma$ of the powerlaw increases. In this work, we found that the spectra become softer with steeper power index as the corona temperature increases during the LHS-HIMS-SIMS transition. This is inconsistent with this framework. Moreover, the inner radius has been stable at the ISCO during the 3 epochs. The behavior of the $\Gamma$ and $kT{\rm{e}}$ is similar to the behavior in GX 339-4 \citep{mot2009} and GRO 1655-40 \citep{joi2008}. 

\begin{figure}
	\includegraphics[width=\columnwidth]{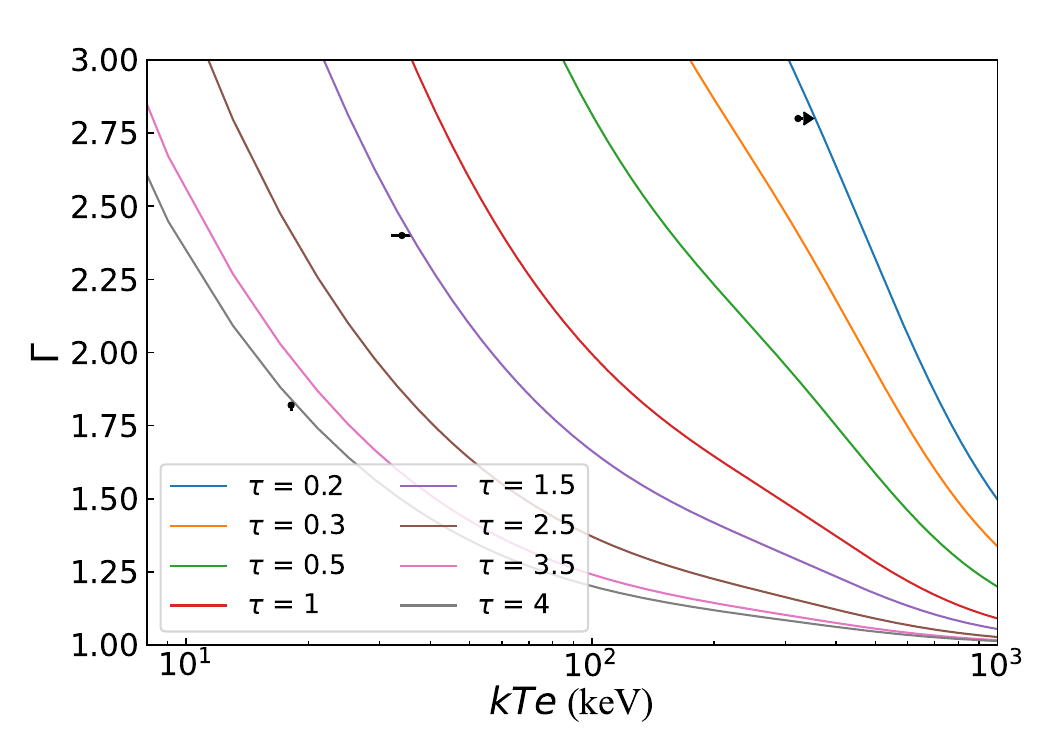}
    \caption{The relationship between $\Gamma$ and $kT_{\rm{e}}$ calculated for the optical depth $\tau$ with value range 0.2-4. The black points are quoted from the best-fitting parameters in M1.}
    \label{fig:gam_kTe}
\end{figure}

Following the above equations, we calculated $\Gamma-kT{\rm{e}}$ plane assuming different values of $\tau$ (Figure \ref{fig:gam_kTe}). The optical depth experienced dramatic change. Its value decreased from $\sim4$ (Epoch 1) to $\sim1.5$ (Epoch 2), then to $\sim0.2$ (Epoch 3). It seems that a dense low-temperature corona in the LHS evolves to a tenuous high-temperature corona after the state transition. The low optical depth of the corona in the SIMS may lead to inefficiency of the Compton scatterings in the corona, of which the cooling is substantially suppressed. Therefore, the corona temperature in Epoch 3 becomes much higher. The behaviour seems to imply that the state transition is accompanied by the coronal evolution. The detailed physical mechanism driving such evolution of the corona is beyond the scope of this work.

\section{Conclusions}
\label{sec:ccl}
The black hole candidate MAXI J1535-571 was caught by \emph{NuSTAR} and \emph{HXMT} at three different states during its 2017 outburst: LHS, HIMS, and SIMS. The results of this work on the broad-band reflection spectra by jointly-fit to these three states found that the inner radius of the disc should be stable at the ISCO. The spin $0.985_{-0.004}^{+0.002}$ indicates a fast rotating black hole in the system. A high inclination angle of the disc is indicated. During the LHS-HIMS transition, the Comptonized component becomes soft, the electron temperature increases slightly, and the thermal component from the disc is relatively comparable. Across the HIMS-SIMS transition, the Comptonized component continues to be soft, the electron temperature shows an abrupt increase, and the thermal component also contributes significantly. We calculated the $\Gamma$-$kT_{\rm{e}}$ panel giving a range value 0.2-4 of the optical depth $\tau$ of the corona. The best-fitting results of $\Gamma$ and $kT_{\rm{e}}$ imply that the $\tau$ varies from $\sim$4 to $\sim$1.5, and to $\sim$0.2 over the state transitions. It is clear that the properties of the corona has changed. The corona evolves from dense low-temperature in the LHS to tenuous high-temperature after the state transition, which seems to imply that the physical properties of the corona has changed during the state transition.

\section*{Acknowledgements}
The authors thank the referee for the helpful comments to improve our manuscript. This work is supported by the NSFC (11773050, 11833007, 12073023), the science research grants
from the China Manned Space Project with NO. CMS-CSST-2021-A06. 
This work made use of data from the NuSTAR mission, a project led by the California Institute of Technology, managed by the Jet Propulsion Laboratory, and funded by the National Aeronautics and Space Administration. This work also made use of the data from the Insight-HXMT mission, a project funded by China National Space Administration (CNSA) and the Chinese Academy of Sciences (CAS). 

\section*{Data availability}
Data used in this article are publicly available from the \emph{NuSTAR} mission (https://heasarc.gsfc.nasa.gov/docs/archive.html) and \emph{HXMT} mission (http://hxmten.ihep.ac.cn).


\bibliographystyle{mnras}
\bibliography{J1535} 



\appendix

\section{Results for individual fits to 3 epochs}

\renewcommand{\arraystretch}{1.2}
\begin{table*}
\centering
\begin{center}
\caption{Best-fitting Parameters of individual fits to 3 epochs}
\begin{tabular}{ccccc}
\toprule
Component&Parameter&Epoch 1&Epoch 2&Epoch 3\\
&&(LHS) &(HIMS) &(SIMS)\\
\midrule
{\tt TBabs}&$N_H$ ($\times10^{22}$ cm$^{-2}$)&6.95$_{-0.19}^{+0.18}$&5.5$_{-0.24}^{+0.26}$&5.33$_{-0.05}^{+0.08}$\\
{\tt diskbb}&$T_{\rm{in}}$ (keV)&0.349$\pm$0.009&0.349$_{-0.016}^{+0.014}$&1.215$_{-0.013}^{+0.006}$\\
&$N_{\rm{disc}}$ ($\times 10^{3}$)&213.88$_{-36.03}^{+41.48}$&230.98$_{-71.02}^{+101.19}$&1.85$\pm$0.04\\
{\tt relxillCp}&$q$&$>$9.81&7.33$_{-0.97}^{+1.44}$&$>$7.22\\
&$i$ (degrees)&77.8$_{-0.69}^{+0.66}$&69.69$_{-2.88}^{+3.07}$&68.88$_{-4.08}^{+5.5}$\\
& $R_{\rm{in}}$ ($R_{\rm{g}}$)&1.35$_{-0.02}^{+0.03}$&1.44$\pm$0.06&$<1.44$\\
& $\Gamma$&1.927$_{-0.016}^{+0.014}$&2.4$_{-0.02}^{+0.021}$&2.777$_{-0.01}^{+0.009}$\\
&log${\rm{\xi}}_{\rm{rel}}$ (erg cm s$^{-1}$)&2.8$_{-0.04}^{+0.05}$&3.11$_{-0.06}^{+0.09}$&3.32$_{-0.06}^{+0.2}$\\
& $kT_{\rm{e}}$ (keV)&25.2$_{-1.59}^{+1.75}$&34.12$_{-3.84}^{+3.68}$&$>$277.43\\
&$R_{\rm{f}}$&1.97$_{-0.27}^{+0.25}$&0.89$\pm$0.21&1.14$_{-0.33}^{+0.24}$\\
&$N_{\rm{rel}}$ ($\times 10^{-2}$)&10.83$\pm$0.16&40.04$_{-1.55}^{+1.79}$&122.11$_{-2.89}^{+3.16}$\\
& $\Gamma_{\rm{HXMT}}$&1.886$_{-0.016}^{+0.015}$&...&...\\
{\tt xillverCp}&log${\rm{\xi}}_{\rm{xil}}$ (erg cm s$^{-1}$)&2.27$_{-0.09}^{+0.15}$&3.68$_{-0.16}^{+0.15}$&...\\
&$N_{\rm{xil}}$ ($\times 10^{-2}$)&1.15$_{-0.15}^{+0.16}$&6.87$_{-1.13}^{+2.28}$&11.88$_{-3.21}^{+3.32}$\\
{\tt constant}&$C_{\rm{FPMB}}$&1.022$\pm$0.001&1.007$_{-0.001}^{+0.002}$&0.997$\pm$0.001\\
&$C_{\rm{LE}}$&0.736$\pm$0.002&0.955$\pm$0.002&0.976$\pm$0.001\\
&$C_{\rm{ME}}$&0.731$_{-0.003}^{+0.004}$&0.96$\pm$0.002&0.936$\pm$0.002\\
&$C_{\rm{HE}}$&0.783$_{-0.012}^{+0.013}$&0.97$\pm$0.015&0.948$_{-0.015}^{+0.016}$\\
\midrule
&$\chi^2$&4576.52&3296.6&3317.74\\
&$\nu$&4273&3323&2991\\
&$\chi^2_{\nu}$&1.07&0.99&1.11\\
\bottomrule
        \end{tabular}
\begin{tablenotes}
        \item \textbf{Notes.} Fitting 3 epochs individually with the model {\tt constant*tbabs(diskbb+relxillCp+xillverCp)}. The spin parameter $a_*$ is fixed at 0.998. The inner radius $R_{\rm{in}}$ is free. The emissivity profile is assumed to be a single powerlaw, for which the emissivity index $q_{\rm{in}}$ = $q_{\rm{out}}$ = $q$. The constant factor is fixed at unity for \emph{NuSTAR}/FPMA, and free for \emph{NuSTAR}/FPMB ($C_{\rm{FPMB}}$), \emph{HXMT}/LE ($C_{\rm{FPMB}}$), ME ($C_{\rm{FPMB}}$), and HE ($C_{\rm{FPMB}}$). The other free parameters listed above: Temperature of the disc ($T_{\rm{in}}$); Inner radius of the disc ($R_{\rm{in}}$); Photon index ($\Gamma$ for and $\Gamma_{\rm{HXMT}}$, which is only different for Epoch 1); Ionization state  ($\mathrm{log}\xi_{\rm{rel}}$ and $\mathrm{log}\xi_{\rm{xil}}$ for {\tt relxillcp} and {\tt xillvercp}, respectively, which is linked for Epoch 3); Electron temperature ($kT_{\rm{e}}$); Reflection fraction ($R_{\rm{f}}$); Normalization constants of {\tt diskbb} ($N_{\rm{disc}}$), {\tt relxillcp} ($N_{\rm{rel}}$) and {\tt xillvercp} ($N_{\rm{xil}}$).
        \end{tablenotes}
\label{tab:FI}
\end{center}
\end{table*}

\bsp	
\label{lastpage}
\end{document}